\documentclass[apj]{emulateapj}

\usepackage{epsfig}
\usepackage{lscape,graphicx,rotating}
\usepackage{apjfonts}

\begin{document}

\title{\Swift\ J2058.4+0516: Discovery of a Possible Second
  Relativistic Tidal Disruption Flare?}

\author{S.~Bradley Cenko\altaffilmark{1},
             Hans A.~Krimm\altaffilmark{2,3},
             Assaf Horesh\altaffilmark{4},
             Arne Rau\altaffilmark{5},
             Dale A.~Frail\altaffilmark{6},             
             Jamie A.~Kennea\altaffilmark{7}, 
             Andrew J.~Levan\altaffilmark{8},
             Stephen T.~Holland\altaffilmark{9},
             Nathaniel R.~Butler\altaffilmark{1,10},
             Robert M.~Quimby\altaffilmark{4},
             Joshua S.~Bloom\altaffilmark{1},
             Alexei V.~Filippenko\altaffilmark{1},
             Avishay Gal-Yam\altaffilmark{11},
             Jochen Greiner\altaffilmark{5},
             S.~R.~Kulkarni\altaffilmark{4},
             Eran O.~Ofek\altaffilmark{4},
             Felipe Olivares E.\altaffilmark{5},
             Patricia Schady\altaffilmark{5},
             Jeffrey M.~Silverman\altaffilmark{1},
             Nial R.~Tanvir\altaffilmark{12},
             and Dong Xu\altaffilmark{11}}
\altaffiltext{1}{Department of Astronomy, University of California,
  Berkeley, CA 94720-3411, USA}
\altaffiltext{2}{CRESST and NASA Goddard Space Flight Center,
  Greenbelt, MD 20771, USA}
\altaffiltext{3}{Universities Space Research Association, 10211
  Wincopin Circle, Suite 500, Columbia, MD 21044, USA}
\altaffiltext{4}{Division of Physics, Mathematics, and Astronomy,
  California Institute of Technology, Pasadena, CA 91125, USA}
\altaffiltext{5}{Max-Planck Institute for Extraterrestrial Physics, 
  Giessenbachstrasse 1, Garching 85748, Germany}
\altaffiltext{6}{National Radio Astronomy Observatory, P.O. Box 0, 
  Socorro, NM 87801, USA}
\altaffiltext{7}{Department of Astronomy \& Astrophysics, The
  Pennslyvania State University, 525 Davey Laboratory, University
  Park, PA 16802, USA}
\altaffiltext{8}{Department of Physics, University of Warwick, 
  Coventry, CV4 7AL, UK}
\altaffiltext{9}{Space Telescope Science Institute, 3700 San Martin
  Drive, Baltimore, MD 21218, USA}
\altaffiltext{10}{Einstein Fellow}
\altaffiltext{11}{Department of Particle Physics and Astrophysics, 
  Faculty of Physics, The Weizmann Institute of Science, Rehovot 
  76100, Israel}
\altaffiltext{12}{Department of Physics and Astronomy, University 
  of Leicester, University Road, Leicester LE1 7RH, UK}

\email{cenko@astro.berkeley.edu}


\shorttitle{\event: A Second Relativistic TDF?}
\shortauthors{Cenko \textit{et al.}}


\newcommand{\Swift}{\textit{Swift}}
\newcommand{\event}{\textit{Sw}~J2058+05}
\newcommand{\eventb}{\textit{Sw}~J1644+57}
\newcommand{\fermi}{\textit{Fermi}}
\newcommand{\mgii}{\ion{Mg}{2} $\lambda \lambda$ 2796, 2803}
\newcommand{\oii}{[\ion{O}{2} $\lambda$ 3727]}
\newcommand{\ip}{\textit{i$^{\prime}$}}
\newcommand{\zp}{\textit{z$^{\prime}$}}
\newcommand{\gp}{\textit{g$^{\prime}$}}
\newcommand{\rp}{\textit{r$^{\prime}$}}
\newcommand{\up}{\textit{u$^{\prime}$}}

\begin{abstract}
We report the discovery by the \Swift\ hard X-ray monitor of
the transient source \Swift\ J2058.4+0516 (\event).  Our
multi-wavelength follow-up campaign uncovered a long-lived (duration
$\gtrsim$ months), luminous X-ray ($L_{\mathrm{X,iso}} \approx
3 \times 10^{47}$\,erg\,s$^{-1}$) and radio ($\nu L_{\nu,\mathrm{iso}} \approx
10^{42}$\,erg\,s$^{-1}$) counterpart.  The associated optical
emission, however, from which we measure a redshift of 1.1853, is
relatively faint, and this is not due to a large amount of dust
extinction in the host galaxy.  Based on numerous
similarities with the recently discovered GRB\,110328A / \Swift\
J164449.3+573451 (\eventb), we suggest that \event\ may be the second 
member of a new class of relativistic outbursts resulting from the tidal
disruption of a star by a supermassive black hole.  If so, the
relative rarity of these sources (compared with the expected rate of 
tidal disruptions) implies that either these outflows
are extremely narrowly collimated ($\theta < 1^{\circ}$), or
only a small fraction of tidal disruptions generate relativistic
ejecta.  Analogous to the case of long-duration gamma-ray bursts
and core-collapse supernovae, we speculate that rapid spin of the 
black hole may be a necessary condition to generate the
relativistic component.  Alternatively, if powered by gas accretion 
(i.e., an active galactic nucleus [AGN]), \event\ would seem to
represent a new mode of variability in these sources, as the observed 
properties appear largely inconsistent with known classes of AGNs
capable of generating relativistic jets (blazars, narrow-line Seyfert
1 galaxies).
\end{abstract}


\section{Introduction}
\label{sec:intro}
The recent discovery of the transient source GRB\,110328A / \Swift\
J164449.3+573451 (\eventb) has unveiled an entirely new class of 
high-energy outbursts \citep{ltc+11,bkg+11,zbs+11}.  Like
long-duration gamma-ray bursts (GRBs), the outburst was believed to
mark the birth
of a relativistic jet, generating luminous X-ray and radio emission.
However, the central engine powering \eventb\ was the
super-massive ($M_{\mathrm{BH}} \lesssim 10^{7}$\, M$_{\odot}$) black
hole in the nucleus of an otherwise normal (i.e., nonactive) galaxy
\citep{bgm+11}.  While not conclusive, the observed emission may 
result from the tidal disruption \citep{r88} of a star (or white dwarf;
\citealt{kp11}) passing too close to the central black hole
\citep{bgm+11,bkg+11,zbs+11,ctl11}.  A number of tidal
disruption flare (TDF) candidates have been previously identified at
longer wavelengths \citep[e.g.,][]{rgd+95,bkd96,kg99,gsz+00,dbe+02,
gmm+06,gbm+08,ghc+09,esf+07,esk+08,mue10,vfg+10,cbk+11}.  But
the robust inference of a newly born relativistic jet, and the
insights into the accretion and jet-formation processes provided
therein, clearly distinguish \eventb\ from previous TDF candidates.

In this work, we report on a recently discovered high-energy
transient, \Swift\ J2058.4+0516 (\event).  Remarkably, \event\
shares many of the properties that made \eventb\ such
an exceptional event, in particular (1) a long-lived (duration
$\gtrsim$ months), super-Eddington ($L_{\mathrm{X,iso}} \approx
3 \times 10^{47}$\,erg\,s$^{-1}$) X-ray outburst; (2) a luminous radio
counterpart, indicating the presence of relativistic
($\Gamma \gtrsim 2$) ejecta; and (3) relatively faint ($M_{U} \approx
-22.7$\,mag) optical emission.  If future observations are
able to establish an astrometric association with the nucleus of the
redshift $z = 1.1853$ host galaxy, and also continue to suggest that this
galaxy does not harbor an active galactic nucleus (AGN), we may have 
identified a second member of this new class of relativistic TDFs.
Alternatively, several classes of AGNs are known to generate
relativistic jets (blazars, narrow-line Seyfert 1 [NLS1] galaxies).
If \event\ is shown to result from an AGN (e.g., via repeated
outbursts over the coming years), we would have uncovered yet another
unique example of the already diverse phenomenology of variability
from active galaxies.

Throughout this work, we adopt a standard $\Lambda$CDM cosmology with
$H_{0}$ = 71\,km s$^{-1}$ Mpc$^{-1}$, $\Omega_{\mathrm{m}} = 0.27$, and
$\Omega_{\Lambda} = 1 - \Omega_{\mathrm{m}} = 0.73$ \citep{sbd+07}.
 All quoted uncertainties are 1$\sigma$ (68\%) confidence
intervals unless otherwise noted, and UT times are used throughout.
Reported magnitudes are in the AB system \citep{og83}.

\begin{figure*}[th]
\plotone{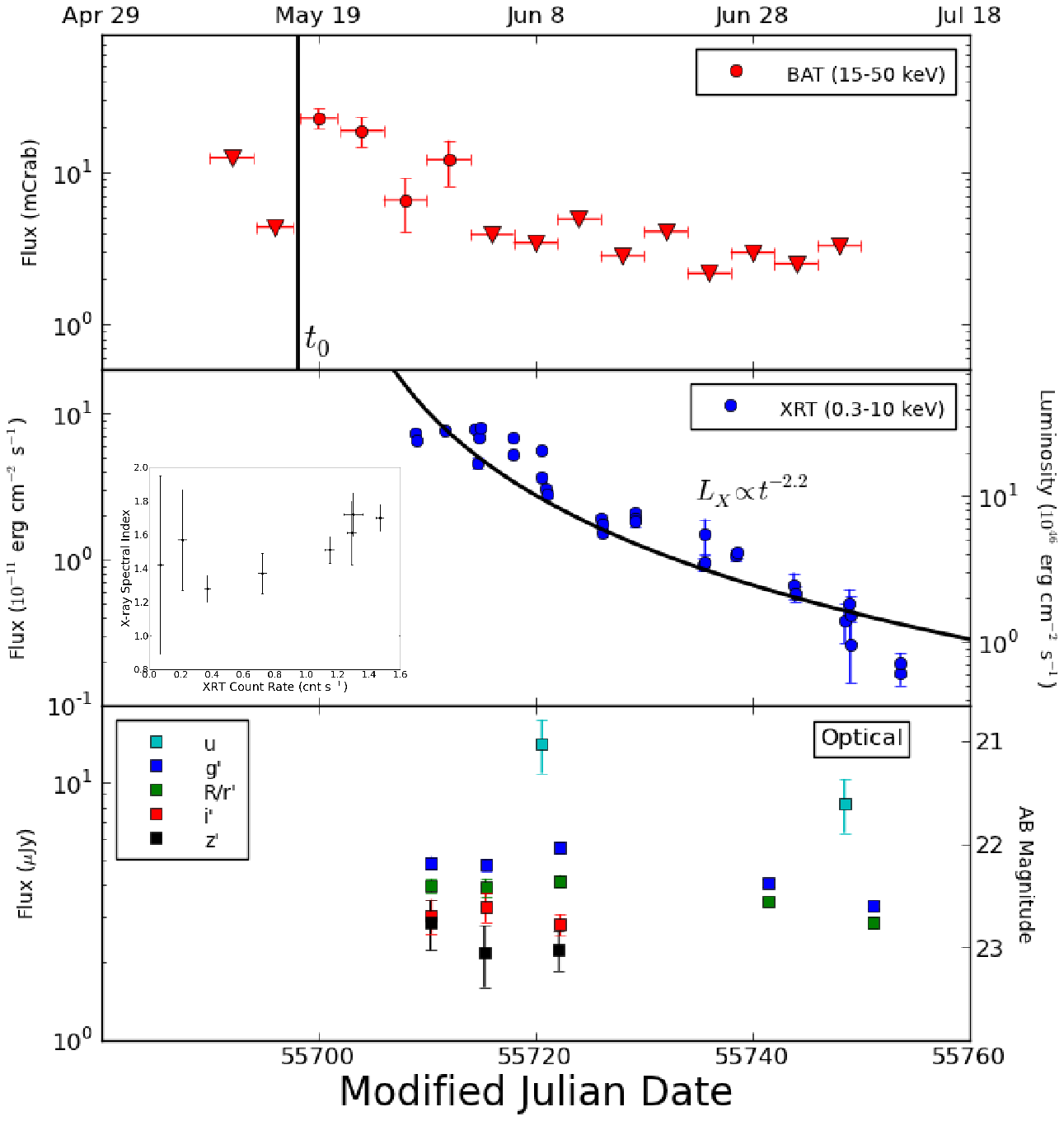}
\caption{Hard X-ray (15--50\,keV), X-ray (0.3--10\,keV), and optical
  light curve of \event.  The inset in the X-ray panels shows the
  derived power-law spectral index ($\Gamma$) as a function of the
  X-ray count rate (i.e., flux).  Inverted triangles represent
  $3\sigma$ upper limits.}
\label{fig:lcurve}
\end{figure*}

\section{Discovery and Observations}
\label{sec:obs}

\subsection{\Swift-BAT Hard X-Ray Monitor Discovery}
\label{sec:bat}
\event\ was discovered by the Burst Alert Telescope (BAT;
\citealt{bbc+05}) on the \Swift\ satellite \citep{gcg+04} as part of
the hard X-ray monitor's automated transient search.  It first reached
the $6\sigma$ discovery threshold in a 4\,d integration covering the
time period 2011 May 17--20, with a 15--50\,keV count rate of $0.0044
\pm 0.0006$\,count\,s$^{-1}$\,cm$^{-2}$ \citep{ATEL.3384}\footnote{Though
  not precisely constrained, we hereafter refer to the discovery time 
  ($t_{0}$) as 00:00 on 2011 May 17 (MJD = 55698).}.  No
significant emission was detected from this location after 2011 June
2.  The resulting BAT (15--50\,keV) light curve is plotted in the top
panel of Figure~\ref{fig:lcurve}.

The observed BAT count rate is too low to measure statistically
significant variations on time scales shorter than 4\,d, or to
constrain the hard X-ray spectrum.  A search of transient monitor 
data on the same 4\,d time scale back to 2005 February shows no 
previous activity from the source with a $3\sigma$ count rate limit 
of $< 0.003$\,count\,s$^{-1}$\,cm$^{-2}$.  

\begin{figure*}
\plotone{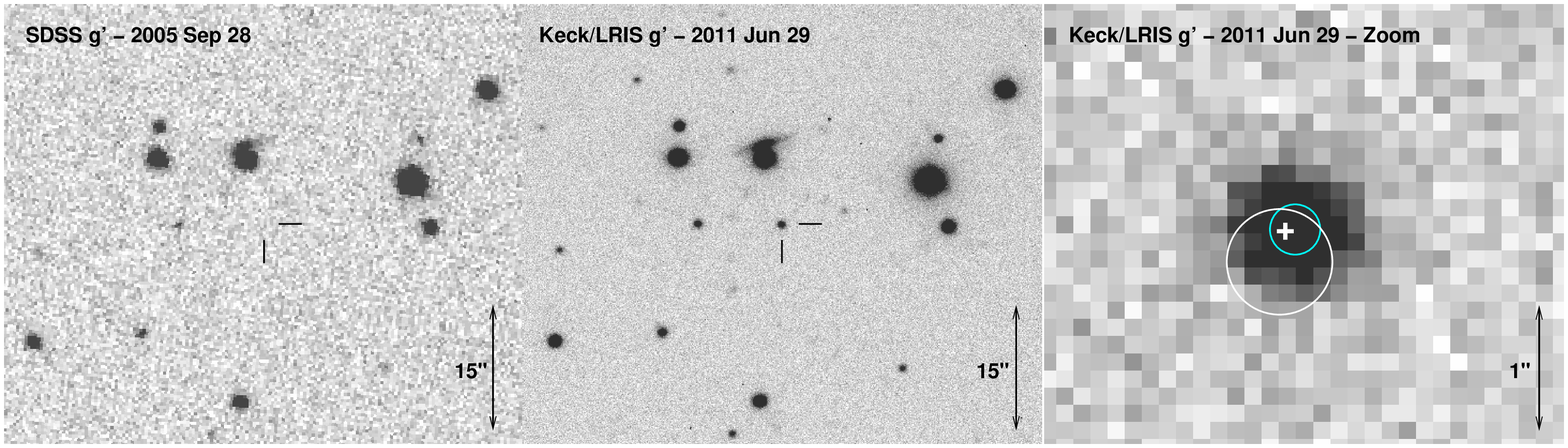}
\caption{Optical finder chart for the field of \event.
  \textit{Left:} Pre-outburst SDSS \gp\ image.  No emission is
  detected at the position of \event\ (indicated with the black
  tick marks) to a limiting magnitude of \gp $> 23.5$\,mag.
  \textit{Middle:} Keck/LRIS \gp\ image of the optical transient.
  The source appears
  point-like in this image (0\farcs7 seeing).  \textit{Right:} Zoom-in
  of the Keck/LRIS image of the transient.  The uncertainty in the
  astrometric tie for the optical frame (aligned with respect to
  2MASS, which is reprojected onto the ICRS reference system)
  is indicated by the cyan circle ($0\farcs2$ radius).  The
  \textit{Chandra}/HRC X-ray localization (based on the native
  \textit{Chandra} astrometry, which is generated with respect to the
  ICRS) is plotted as the white circle ($0\farcs4$ radius), while the
  EVLA radio position (50\,mas uncertainty) is shown as the
  white cross.  All three positions are entirely consistent.
  All images are oriented with north up and east to the left.}
\label{fig:finder}
\end{figure*}


\subsection{X-rays}
\label{sec:xray}
To confirm this discovery, we requested a \Swift\
target-of-opportunity (ToO) observation, which began at 21:56 
on 2011 May 28.  Data obtained by the X-ray Telescope (XRT; 
\citealt{bhn+05}) were reduced using the online analysis 
tools of \citet{ebp+09}, and XRT spectra were fitted using 
{\ttfamily xspec12.6}\footnote{See
http://heasarc.gsfc.nasa.gov/docs/xanadu/xspec/.}.  After correcting
the XRT astrometry following \citet{gtb+98}, we identified a bright
point source at (J2000.0) coordinates $\alpha = 20^{\mathrm{h}} 
58^{\mathrm{m}} 19.85^{\mathrm{s}}$, $\delta = +05^{\circ} 13\arcmin 
33\farcs0$, with a 90\% localization radius of $1\farcs7$.  The XRT
continued to monitor \event; the resulting 0.3--10\,keV light curve 
is plotted in the middle panel of Figure~\ref{fig:lcurve}.

Superimposed on the secular decline (reasonably well described by a
power law: $L_{\rm X} \propto t^{-2.2}$), the X-ray light curve shows
some degree of variability on relatively short time scales (compared
to the time since discovery).  The most significant flaring is
detected on 2011 June 2, with a change in flux of a factor of 1.5 
occurring on a time scale of $\delta t \approx 10^{4}$\,s.  No 
significant variability is detected on shorter time scales; however, 
given the signal-to-noise ratio (SNR), we are not sensitive to similar 
variations (i.e, factor of 2) on time scales shorter than $\sim 10^{3}$\,s.  

The average photon-counting mode spectrum is reasonably well described
by an absorbed power law with photon index $\Gamma = 1.61 \pm 0.12$
and $N_{\mathrm{H,host}} = (2.6 \pm 1.6) \times 10^{21}$\,cm$^{-2}$
($\chi^{2} = 65.95$ for 54 degrees of freedom, d.o.f.; see
\citealt{bk07} for details of the spectral analysis).  Using this
time-averaged spectrum, we find that the peak (unabsorbed) 0.3--10\,keV
flux is $f_{\rm X} \approx 7.9 \times
10^{-11}$\,erg\,cm$^{-2}$\,s$^{-1}$.  The
requirement for absorption in excess of the Galactic value
($N_{\mathrm{H,Gal}} = 6.5 \times 10^{20}$; \citealt{kbh+05}) is only
significant at the $3.2\sigma$ level.  Interestingly, the
time-resolved hardness ratio is inversely correlated with the source
flux (Figure~\ref{fig:lcurve}), a behavior commonly seen in Galactic 
black hole binaries (e.g., \citealt{rm06}).  

To improve the astrometric precision of the X-ray localization, we
obtained ToO observations of \event\ with the High Resolution
Camera (HRC; \citealt{mck+97}) on the \textit{Chandra} X-ray
Observatory.  Observations began at 14:51 on 2011 June 22, for a total
live exposure of 5.2\,ks.  \event\ is well detected with
(J2000.0) coordinates $\alpha = 20^{\mathrm{h}} 58^{\mathrm{m}}
19.90^{\mathrm{s}}$, $\delta = +05^{\circ} 13\arcmin 32\farcs0$, and
an astrometric uncertainty of $0\farcs4$ (Figure~\ref{fig:finder}; 
the lack of additional point sources in the HRC image precludes us
from improving the native \textit{Chandra} astrometric precision).  
The X-ray flux shows weak evidence ($1.5\sigma$) for a decline 
over the course of the entire observation, but no statistically 
significant variability on shorter time scales.

Finally, we note that the location of \event\ was observed 
on 2000 March 16 by the Position Sensitive Proportional Counters 
(PSPC) onboard \textit{ROSAT} as part of the All-Sky Survey 
\citep{vab+99}.  No sources are detected in the field to a limit of 
$f_{\rm X}$(0.1--2.4\,keV) $< 10^{-13}$\,erg\,cm$^{-2}$\,s$^{-1}$, several
orders of magnitude fainter than the observed outburst.

\subsection{UV/Optical/NIR Photometry}
\label{sec:obs:opt}
Following the discovery of the X-ray counterpart, we initiated a
campaign to observe \event\ in the ultraviolet (UV), optical,
and near-infrared (NIR), with the \Swift\ Ultraviolet-Optical
Telescope (UVOT; \citealt{rkm+05}), the 7-channel imager GROND
\citep{gbc+08} mounted on the 2.2\,m telescope at La Silla
Observatory, the Low Resolution Imaging Spectrometer (LRIS;
\citealt{occ+95}) mounted on the 10\,m Keck I telescope, the
Wide-Field Camera (WFCAM) on the 
United Kingdom Infrared Telescope (UKIRT), and the Auxiliary-port
Camera (ACAM) on the 4.2\,m William Herschel Telescope.  All
observations were reduced following standard procedures and 
photometrically calibrated with respect to the 2 Micron All Sky Survey 
(2MASS; \citealt{scs+06}) in the NIR, Sloan Digital Sky Survey (SDSS; 
\citealt{aaa+09e}) in the optical, and following \citet{pbp+08} for 
the UVOT.  

The optical counterpart of \event\ was first identified in our
GROND imaging obtained on 2011 May 29 at (J2000.0) coordinates $\alpha
= 20^{\mathrm{h}} 58^{\mathrm{m}} 19.90^{\mathrm{s}}$, $\delta =
+05^{\circ} 13\arcmin 32\farcs2$, with an astrometric uncertainty of
0\farcs2 (radius; \citealt{ATEL.3390,ATEL.3425}).  The
counterpart appears point-like in all our images, with the tightest
constraints provided by our Keck/LRIS images on 2011 Jun 29 (0\farcs7
seeing; Figure~\ref{fig:finder}).  Since discovery, we detect
statistically significant fading in the bluer filters, but the degree
of optical variability is much smaller than that observed in the
X-rays.  A full listing of our photometry is provided in
Table~\ref{tab:opt}, while the light curves are plotted in the bottom 
panel of Figure~\ref{fig:lcurve}.

Pre-outburst imaging of the location of \event\ was 
obtained on 2005 September 28 as part of SDSS. 
No source consistent with the location of 
\event\ appears in the SDSS catalogs.  Manually inspecting the 
images, we calculate $3\sigma$ upper limits of $\up > 21.9$, 
$\gp > 23.5$, $\rp > 23.5$, $\ip > 23.0$, and
$\zp > 21.7$\,mag.  The lack of a pre-existing counterpart in SDSS is
indicative that the observed optical emission is dominated by
transient light and not any underlying host galaxy.


\subsection{Optical Spectroscopy}
We obtained a series of optical spectra of \event\ on 2011
June 1 with the Deep Imaging Multi-Object Spectrograph (DEIMOS;
\citealt{fpk+03}) mounted on the 10\,m Keck II telescope.  The
instrument was configured with the 600\,lines\,mm$^{-1}$ grating,
providing spectral coverage over the region $\lambda =
4500$--9500\,\AA\ with a spectral resolution of 3.5\,\AA.  The spectra
were optimally extracted \citep{h86}, and the rectification and sky
subtraction were performed following the procedure described by
\citet{k03}.  The slit was oriented at the parallactic angle to
minimize losses due to atmospheric dispersion
\citep{f82}.

\begin{figure*}
\plotone{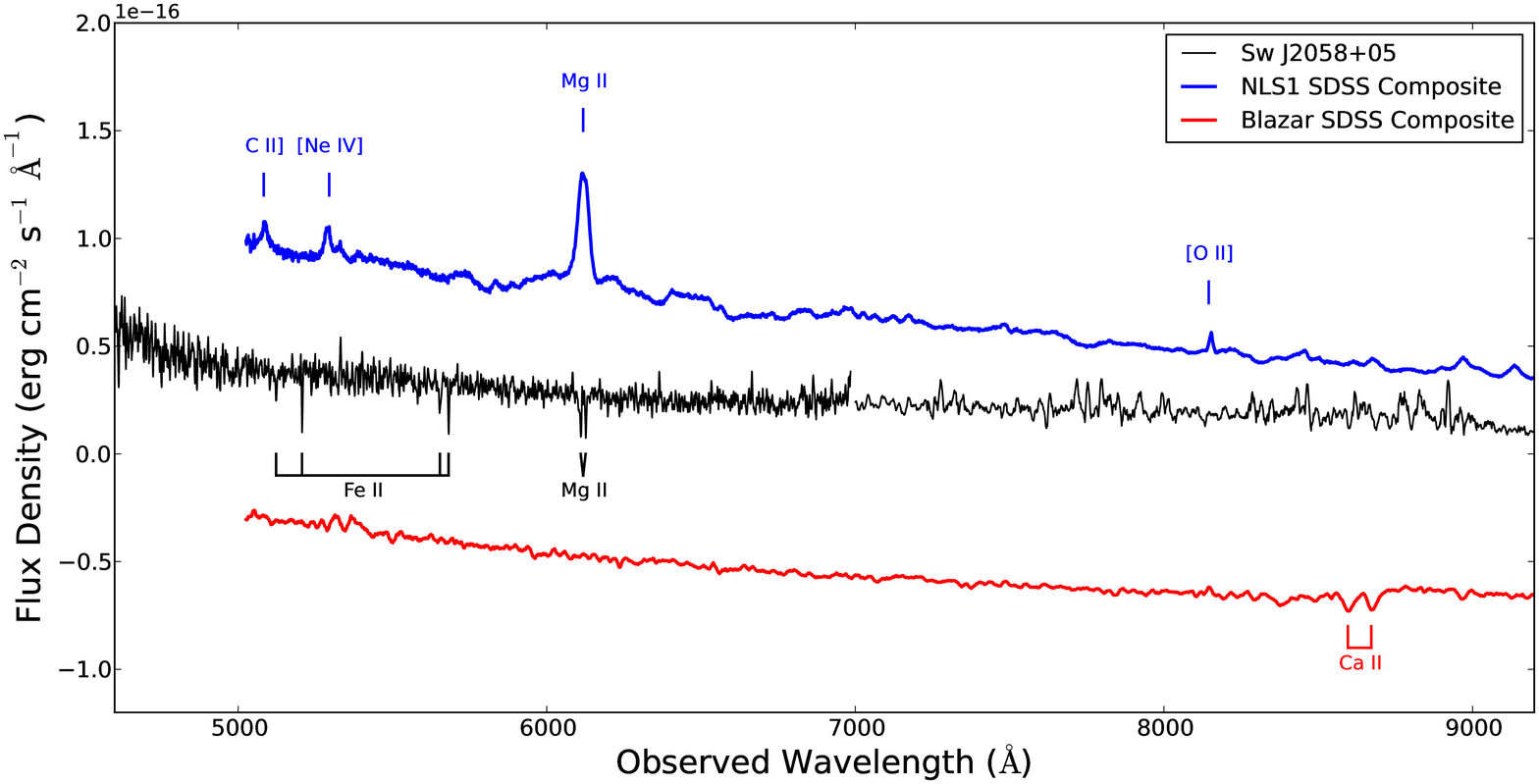}
\caption{Keck/DEIMOS optical spectrum of \event\  (black).  The significant
  absorption features, corresponding to \ion{Mg}{2} $\lambda
  \lambda$2796, 2803, \ion{Fe}{2} $\lambda$2600, \ion{Fe}{2}
  $\lambda$2587, \ion{Fe}{2} $\lambda$2383, and \ion{Fe}{2}
  $\lambda$2344 at a common redshift of $1.1853 \pm
  0.0004$, are marked.  Also plotted is an average of the
  478 known NLS1 galaxies (blue) and 68 blazars (red) at
  $z > 0.5$ with spectra in SDSS .  All but one of these NLS1 galaxies
  exhibit broad \ion{Mg}{2} emission (as well as weaker \ion{C}{2}],
  [\ion{Ne}{4}], and [\ion{O}{2}]), which is entirely lacking from
  \event.  While the blazars (by definition) lack strong emission
  features, most exhibit \ion{Ca}{2} $\lambda \lambda$3934, 3968
  absorption, and none have \ion{Mg}{2} detected in absorption.
  Note that the red-side spectrum of \event (longward of
  $\sim 7000$\,\AA) suffers from relatively poor night-sky
  subtraction, giving rise to apparent emission lines. Also, the weak
  emission features near 5300\,\AA\ in the blazar spectrum are
  likely to be noise.}
\label{fig:spec}
\end{figure*}

A portion of the resulting spectrum is plotted in
Figure~\ref{fig:spec}.  Superimposed on a relatively blue continuum,
we identify several strong (rest-frame equivalent width
$W_{\mathrm{r}} \gtrsim 1$\,\AA) absorption features corresponding to
\ion{Mg}{2} $\lambda\lambda$2796, 2803, \ion{Fe}{2} $\lambda$2600, 
\ion{Fe}{2} $\lambda$2587, \ion{Fe}{2} $\lambda$2383, and
\ion{Fe}{2} $\lambda$2344 at $z = 1.1853 \pm
0.0004$.  Given the $U$-band detection ($z \lesssim 2$ assuming 
$\lambda_{\mathrm{Ly}\alpha} \lesssim \lambda_{U}$), we assume this 
redshift corresponds to the host galaxy of \event.  No other significant
features are detected, either in emission or absorption, over the 
observed bandpass.  Specifically, we limit the flux from [\ion{O}{2}] 
$\lambda$3727 to be $f < 3 \times 10^{-17}$\,erg\,cm$^{-2}$\,s$^{-1}$
($3\sigma$).  

A second spectrum was obtained with Keck/LRIS on 2011
June 29, covering the wavelength range
3500--9500\,\AA.  We find no significant evolution in either the
shape of the continuum or the observed absorption features over this
time period.

\subsection{Radio}
\label{sec:radio}
We observed the location of \event\ with the NRAO Expanded
Very Large Array (EVLA; \citealt{pcb+11})\footnote{The 
National Radio Astronomy Observatory is a facility 
of the National Science Foundation operated under cooperative 
agreement by Associated Universities, Inc.} on 2011 June 29.36
and 2011 July 19.39. Observations were conducted in the C band
(4--8\,GHz) on June 29 and in the K (8--12\,GHz) and X (18--26\,GHz)
bands on July 19.  
The array was in the A configuration for both epochs, and data were 
processed using standard routines in the AIPS environment.

Within the EVLA field of view, we detect a single point source with
(J2000.0) coordinates $\alpha = 20^{\mathrm{h}} 58^{\mathrm{m}}
19.898^{\mathrm{s}}$, $\delta = +05^{\circ} 13\arcmin
32\farcs25$, with an (absolute) astrometric uncertainty of 50\,mas.  
We report the following flux densities: on June 29, $f_{\nu} (\nu = 
4.5$\,GHz$) = 0.88 \pm 0.05$\,mJy and $f_{\nu}(\nu = 
7.9$\,GHz$) = 0.84 \pm 0.04$\,mJy; on July 19, $f_{\nu} (\nu =
8.4$\,GHz$) = 1.34 \pm 0.04$\,mJy and $f_{\nu}(\nu = 22.5$\,GHz$) =
1.21 \pm 0.15$\,mJy.

The EVLA localization is plotted as a cross in the right panel of
Figure~\ref{fig:finder}.  Native EVLA astrometry is relative to the
ICRS system, which we have also used for the optical (via the 2MASS
point source catalog) and X-ray localizations (via the native
\textit{Chandra} pointing solution).  All three positions are 
coincident, implying that the emission we are observing in different 
bandpasses all results from the same physical source.

This field had been observed previously by the VLA at 1.4\,GHz, once on
1996 July 8 as part of the NRAO VLA Sky Survey (NVVS;
\citealt{ccg+98}), and three times in 2009 (March 9, March 16, and May
9) as part of the Faint Images of the Radio Sky at Twenty-Centimeters 
(FIRST; \citealt{bwh95}) survey. We inspected the images from these 
surveys and found no source at the position of \event, with 
3$\sigma$ limits of $f_{\nu} < 1.5$\,mJy (NVSS) and 0.38\,mJy (FIRST).

\section{Discussion}
\label{sec:disc}

\subsection{Comparison to \eventb\ and Known Classes of Active Galaxies}
\label{sec:swj1644}
\begin{figure*}
\plotone{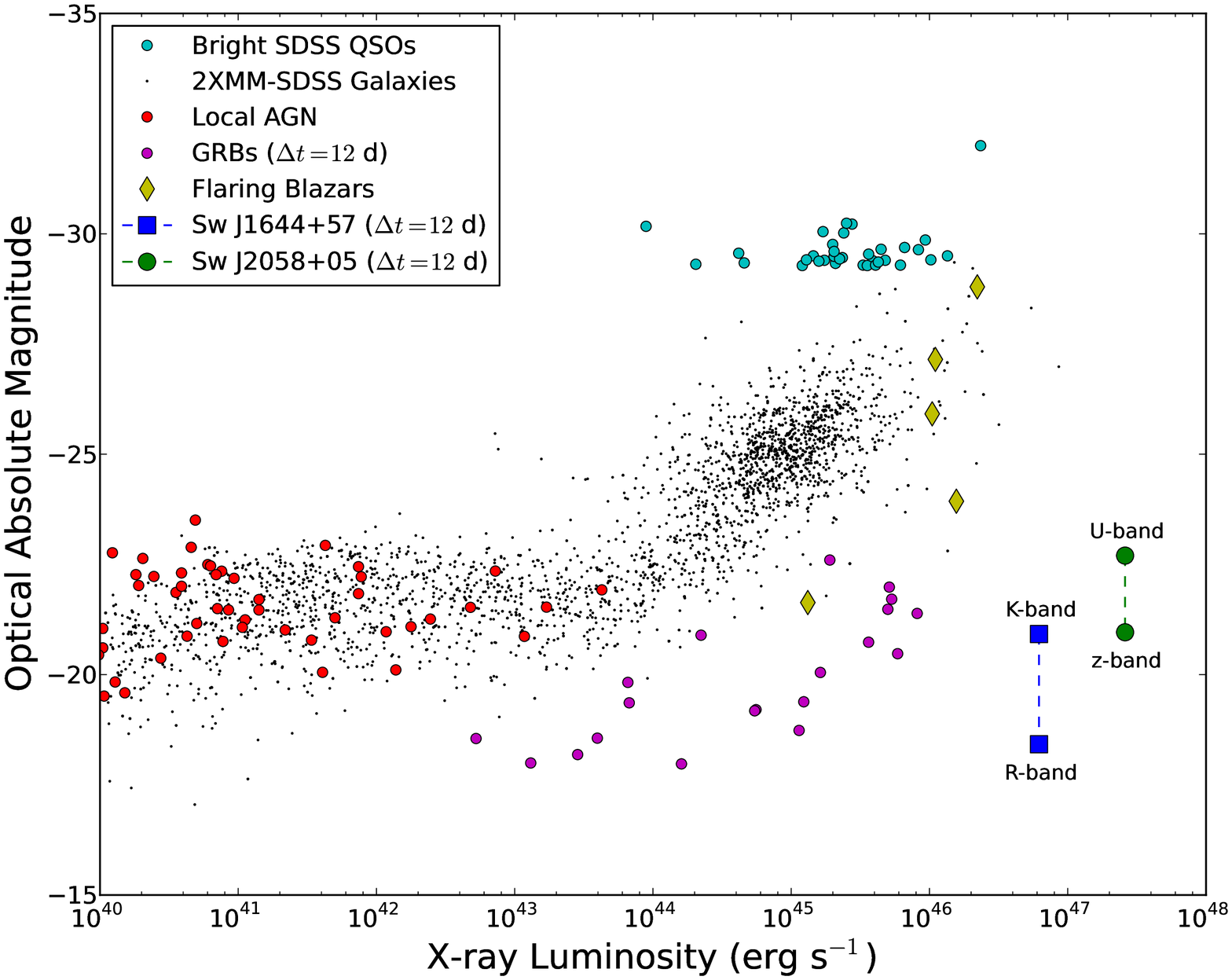}
\caption{X-ray and optical luminosity of \event, compared with
  a sample of nearby active galaxies (red circles; \citealt{h09}),
  luminous quasars from SDSS (cyan circles), galaxies from SDSS with
  XMM counterparts (black dots; \citealt{pmc+11}), long-duration
  GRB afterglows (extrapolated to a common epoch of $\Delta t = 12$\,d
  after the GRB trigger; \citealt{kkz+07,ebp+09}), and high-energy
  outbursts from known blazars (Markarian\,501, \citealt{pvt+98};
  PKS\,2155-304, \citealt{fgt+07}; PKS\,0537-441, \citealt{prt+07};
  3C\,279, \citealt{wpu+98}; Markarian\,421, \citealt{fbb+08}). Compared with
  these sources (and at the same epoch post-discovery as the GRBs),
  both \event\ and \eventb\ exhibit extremely luminous X-ray emission,
  yet relatively faint optical emission.  While for \eventb\ this is
  to some extent due to dust extinction in the host galaxy (for which we
  have not corrected here), the blue
  UV/optical SED of \event\ suggests at most a modest host-galaxy column
  density.  We note that we have not K-corrected the absolute magnitudes
  reported here (aside from cosmological stretch) due to the
  relatively uncertain intrinsic SEDs.  Adapted from \citet{ltc+11}.}
\label{fig:lxlo}
\end{figure*}

In Figure~\ref{fig:lxlo}, we plot the X-ray and optical luminosity of
\event\ at an (observer-frame) time of $\Delta t \approx 12$\,d after
discovery.  For comparison, we also plot analogous measurements for a
sample of long-duration GRB X-ray afterglows (at the same epoch
post-burst), as well as nearby AGNs, more 
distant, luminous quasars from SDSS, and some of the most dramatically
variable blazars \textit{observed while in outburst}.  With an isotropic X-ray 
luminosity $L_{\mathrm{X,iso}} \approx 3 \times
10^{47}$\,erg\,s$^{-1}$, \event\ is much too luminous at this
late time to result from a GRB.  Yet with an absolute magnitude of 
$M_{U} \approx -22.7$\,mag, the observed optical
emission is several orders of magnitude underluminous compared with
the brightest X-ray quasars and blazars.  Even compared with simultaneous X-ray
and optical observations of the strongest high-energy flares from
some of the best-known blazars (e.g., Markarian\,501, \citealt{pvt+98};
3C\,279, \citealt{wpu+98}; Markarian\,421, \citealt{fbb+08}), the ratio
of X-ray to optical luminosity appears to be quite unique. However,
\event\ occupies a nearly identical region of this phase space as \eventb.

\begin{figure*}
\plotone{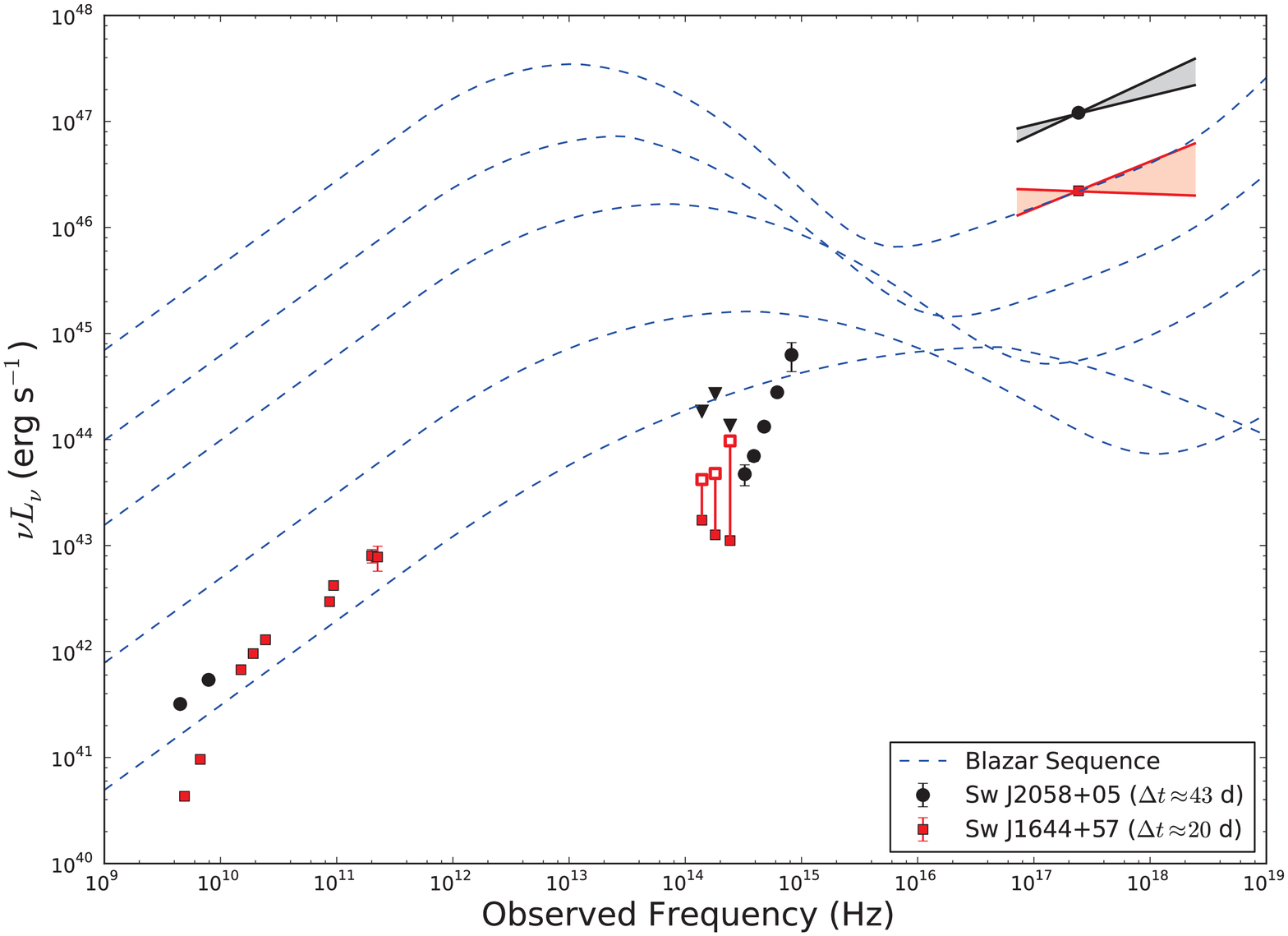}
\caption{SEDs of \event\ and \eventb,
  compared with the blazar ``sequence'' from
 \citet{fmc+98}.  The open squares attempt to correct for the large
 (but uncertain) host-galaxy extinction in the case of \eventb\ by
 assuming $A_{V} = 5$\,mag.  The large X-ray luminosities of both
 \event\ and \eventb\ are incompatible with the observed optical and
 radio fluxes, which would imply an intrinsically fainter source.
 Both objects have complex SEDs that are difficult to reconcile with
 only a single emission component (e.g., synchrotron).}
\label{fig:sed}
\end{figure*}

This similarity is further reinforced when adding the radio
observations to the broadband spectral energy distribution (SED).  In
Figure~\ref{fig:sed} we plot the SED of \event\ at $\Delta t \approx 
43$\,d after discovery, compared with analogous measurements for a 
sample of blazars, the most dramatically variable class of AGNs (and
also known sources of relativistic jets; \citealt{up95}; see 
\S\ref{sec:physics}).  Blazars
exhibit a well-defined luminosity ``sequence'' whereby the 
lower-frequency maximum of their double-peaked SED occurs at lower 
energies for more luminous sources \citep{fmc+98}.  While the 
radio and optical emission from \event\ are comparable to those seen 
in low-luminosity blazars, the X-ray emission outshines even the 
brightest sources of this class.  On the other hand, the SED of
\eventb, at a time of $\Delta t \approx 20$\,d, provides a good 
match to the observed properties of \event\ (particularly after
correcting for the large but uncertain dust extinction from \eventb).

While the sequence plotted in Figure~\ref{fig:sed} may
represent the ensemble properties of blazars as a class, individual
sources may exhibit quite different behavior.  During an
outburst, blazar SEDs can undergo radical changes --- for example,
as a result of a rapid brightening in 1997, the low-energy (synchrotron)
spectral peak in Markarian\,501 shifted frequencies by approximately
two orders of magnitude \citep{pvt+98}.  We therefore also compare the
observed SED of \event\ with that of several high-energy outbursts from known
blazars in Figure~\ref{fig:xrayopt}.  Even when compared with these
extreme cases, the ratio of the X-ray to optical flux observed in
\event\ stands out, being similar only to the broadband properties
of \eventb.  

The sole remaining defining characteristic of \eventb\ is its
association with the nucleus of a nonactive galaxy
\citep{ltc+11,zbs+11}.  At the current time, the observed optical
emission from \event\ appears to be dominated by the transient,
and we have no constraints on the location of the host nucleus from
pre-outburst observations.  Future high-resolution observations, when
the transient light has faded, should be able to constrain the
transient-nucleus offset.  That said, the detection of strong 
absorption features indicates that the line of sight penetrates a dense 
region of the host galaxy (as in the case of some GRBs).  

Finally, we utilize the optical spectrum to search for evidence of
past AGN activity.  Like blazars, NLS1 galaxies are
another subclass of AGNs that can sometimes power relativistic jets
(e.g., \citealt{aaa+09g,aaa+09h,fgk+11}; \S\ref{sec:physics}).  Similarly, 
some NLS1 galaxies have been associated with luminous ($L_{\rm X} \approx
10^{44}$\,erg\,s$^{-1}$) X-ray outbursts (e.g.,
\citealt{gbm+95,bpf95}).  In the case of
\eventb, all of the detected emission lines were unresolved, and standard
diagnostic diagrams revealed no evidence for ionization from a hidden
(i.e., Compton thick) AGN \citep{ltc+11,zbs+11}.  The larger redshift
precludes such an analysis for \event, as the standard diagnostic
lines do not fall in the optical bandpass.

However, we have compiled all of
the available spectra from SDSS of known NLS1 galaxies (from the
catalog of \citealt{vv10}) with $z > 0.5$ (478 objects), and used
these to create an NLS1 template.  This composite spectrum
is plotted alongside that of \event\ in Figure~\ref{fig:spec}.  
The template exhibits strong, broad
emission from \ion{C}{2}] $\lambda$2324, [\ion{Ne}{4}] $\lambda 
\lambda$2422, 2424, and in particular \ion{Mg}{2} $\lambda 
\lambda$2796, 2803.  None of these emission lines are detected in our
spectrum of \event.  In fact, of the 478 known NLS1 galaxies in SDSS
with $z > 0.5$, all but one have well-detected ($> 3\sigma$)
\ion{Mg}{2} in emission.  Unless exceedingly rare, the optical
spectrum of \event\ suggests that it is not an NLS1 galaxy.   

We have furthermore performed a similar analysis with all known BL Lac
sources from \citet{vv10} at $z > 0.5$ in the SDSS archive (68 sources);
the resulting median spectrum is shown in red in Figure~\ref{fig:spec}.
By definition, BL Lacs lack the bright emission features present in
the spectra of NLS1 galaxies, and therefore they more closely
resemble the observed spectrum of \event.  However, BL Lac spectra do
typically exhibit \ion{Ca}{2} $\lambda \lambda$3934, 3968 in
absorption, which is lacking in \event, and not a single BL Lac in the
SDSS database exhibits strong \ion{Mg}{2} absorption.  While we cannot
entirely rule out a new mode of variability in a blazar, the unique
SED and optical spectrum distinguish \event\
from known members of this class as well.

\subsection{Basic Physical Properties}
\label{sec:physics}
We have shown in the previous section that \event\ does not appear
consistent with any known class of AGNs (particularly blazars and
NLS1).  Based on the many
similarities with \eventb,  we can use the same arguments to derive the
basic physical parameters for \event\ (e.g., \citealt{bgm+11}).
From our NIR observations, we derive an upper limit on the 
host-galaxy absolute $V$-band magnitude of $M_{V} \gtrsim -21$\,mag (we
have used an S0 galaxy template from \citealt{kcb+96} for the
K-correction based on the lack of emission features in 
the optical spectrum).  Applying the black hole mass vs. bulge 
luminosity \citep{mtr98} correlation from \citet{lfr+07} and assuming 
a bulge-dominated system, we limit the mass of the putative central black 
hole to be $M_{\mathrm{BH}} \lesssim 2 \times 10^{8}$\, M$_{\odot}$.

Separately, the shortest observed X-ray variability time scale can
be used to constrain the central black hole mass using causality
arguments (e.g., \citealt{GCN.11851}).  For an observer-frame
variability time scale of $\delta t \lesssim 10^{4}$\,s, we find
$M_{\mathrm{BH}} \lesssim 5 \times 10^{8}$\, M$_{\odot}$.  The most
dramatic short time scale variability observed from \eventb\ occurred
at $\Delta t \lesssim 10$\,d \citep{bgm+11}, before we commenced X-ray
observations of \event.  But even at late times, the X-ray light curve
of \event\ exhibits a degree of variability (changes of an order of
magnitude on time scales of a few days) not seen in \event,
suggesting an imperfect analogy between these two sources.

Finally, we can constrain the central mass using the black hole
fundamental plane, a relationship between mass, X-ray luminosity, and radio
luminosity (e.g., \citealt{gcm+09}).  Application of this relation requires
knowledge of the beaming fraction, for which we adopt a value of
$f_{b} \equiv (1 - \cos \theta) \approx 10^{-2}$ (see below).
Following \citet{mg11}, we find $M_{\mathrm{BH}} \approx 5 \times
10^{7}$\, M$_{\odot}$; however, we caution that the scatter in this 
relationship is even larger than that found in the black hole mass 
vs. bulge luminosity relation.

The derived black hole mass, $M_{\mathrm{BH}} \lesssim
10^{8}$\, M$_{\odot}$, is of vital importance for two main reasons.
First, the tidal disruption of a solar-type star will only occur outside the
event horizon for $M_{\mathrm{BH}} \lesssim 2 \times
10^{8}$\, M$_{\odot}$.  As such, if the origin of the accreting
material is indeed the tidal disruption of a nondegenerate star 
(or more massive
object), the mass of the central black hole in the host galaxy of 
\event\ must be sufficiently small\footnote{The limits on the central 
black hole of the host galaxy of \eventb\ are even more stringent
($M \lesssim 10^{7}$\, M$_{\odot}$; \citealt{ltc+11,bkg+11}), and may 
even allow for the disruption of a white dwarf \citep{kp11}.}.  Future
observations of the host, when the optical transient light has faded,
should provide a significantly improved estimate of $M_{\mathrm{BH}}$.

Second, for $M_{\mathrm{BH}} \lesssim 10^{8}$\, M$_{\odot}$, the
observed X-ray emission exceeds the Eddington luminosity.  In
particular, the peak isotropic 0.3--10\,keV X-ray luminosity of 
$L_{\mathrm{X,iso}} \approx 3 \times 10^{47}$\,erg\,s$^{-1}$ exceeds the 
Eddington value by more than an order of magnitude.  Given the long-lived 
nature of the central engine, this suggests the outflow is likely to be 
collimated with a beaming factor $f_{b} \lesssim 10^{-1}$ (cf., 
\citealt{s11}).  An even larger degree of collimation is required to 
reconcile the observed rate of these outbursts with theoretical 
predictions for TDFs (\S\ref{sec:implications}).

More robust evidence for beaming is provided by the large radio
luminosity.  It is well known that the brightness temperature of an
incoherent radio source cannot exceed $10^{11}$\,K for extended 
periods of time \citep{r94,kfw+98}.  Utilizing this fact, we can
derive a lower limit on the angular size of the radio-emitting region, 
$\theta \gtrsim 14$\,$\mu$as.  Assuming that the outflow commencement
coincides with the hard X-ray discovery, this implies a mean 
expansion velocity of $\beta \equiv v / c \geq 0.88$, or a mildly
relativistic expansion with $\Gamma \geq 2.1$.  Furthermore, the radio emission
will be collimated into a viewing angle $\theta \lesssim 1 / \Gamma$
due to relativistic beaming effects.

If we integrate over the X-ray light curve to date (2011 July 20), we find
that \event\ has emitted a total 0.3--10\,keV fluence of $S_{\rm X}
\approx 1.0 \times 10^{-4}$\,erg\,cm$^{-2}$.  If we assume this
accounts for $\sim 1/3$ of the broadband luminosity, the total
isotropic energy release is $E_{\mathrm{iso}} \approx 10^{54}$\,erg,
comparable to that of \eventb\ at a similar epoch.  For a typical radiative 
efficiency for accretion power ($\eta \equiv E_{\mathrm{rad}} / M 
c^{2} \lesssim 0.1$), the tidal disruption of a solar-type star would 
require a significant beaming correction.  If only a small fraction of 
the available energy is allocated to the observed relativistic 
component ($\lesssim 1$\%), this would imply $f_{b} \lesssim 10^{-3}$.  

\begin{figure*}
\plotone{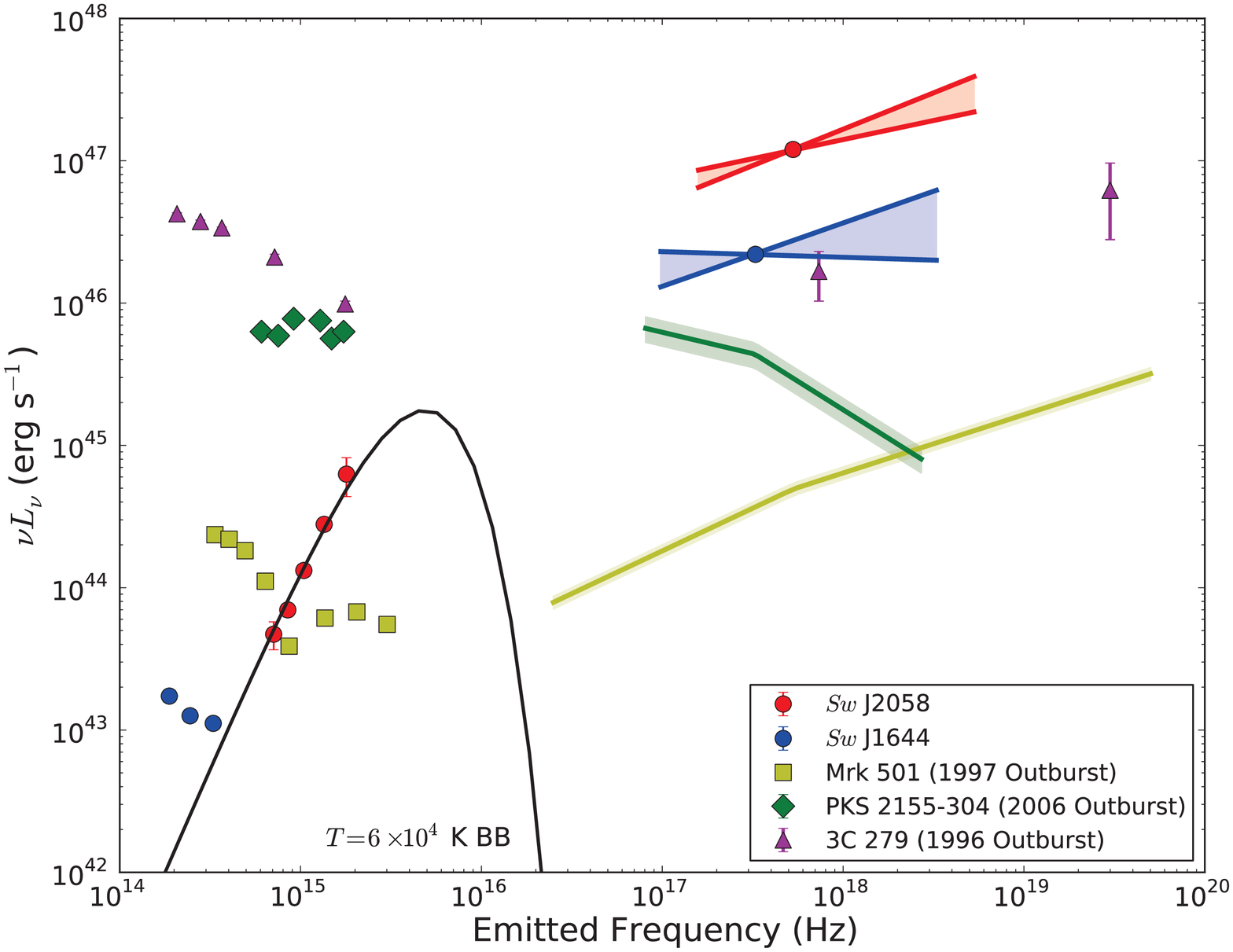}
\caption{Optical and X-ray SEDs of \event\
  and \eventb, compared with some of the most dramatically variable
  blazars \textit{while in outburst} (Markarian\,501,
  \citealt{pvt+98}; PKS\,2155-304, \citealt{fgt+07}; 3C\,279,
  \citealt{wpu+98}).  Unlike any of the other objects plotted here,
  the rest-frame UV SED of \event\ is well represented by a blackbody
  by $T \gtrsim 6 \times 10^{4}$\,K (solid black line).}
\label{fig:xrayopt}
\end{figure*}

In addition to the weaker degree of X-ray variability, we wish to draw 
attention to two other important differences between these sources.  
Unlike that case of \eventb, the observed UV/optical SED
of \event\ is quite blue.  In fact, the rest-frame UV data for
\event\ can be well fit by a blackbody spectrum
(Fig.~\ref{fig:xrayopt}), with $T_{\mathrm{BB}} \gtrsim 6 \times
10^{4}$\,K, $L_{\mathrm{BB}} \gtrsim 10^{45}$\,erg\,s$^{-1}$, and
$R_{\mathrm{BB}} \gtrsim 10$\,AU.\footnote{Because the observed filters
  fall on the Rayleigh-Jeans tail, our constraints on the blackbody
  temperature, luminosity, and radius are essentially lower limits.}
The derived blackbody parameters are similar to the thermal
emission observed from previous TDF candidates
discovered at UV and optical wavelengths 
\citep{gmm+06,gbm+08,ghc+09,vfg+10,cbk+11}, and are strongly
suggestive of an accretion disk origin for the optical light.  In this
sense, \event\ may serve as a link between the bright, nonthermal
X-rays observed in the relativistic TDF candidates,
and the thermal disk emission from ``classical'' TDFs.  Such
comparisons were not possible for \eventb\ due to dust in the host
galaxy.

More importantly, the radio spectral index of \event\ appears to be quite
flat ($f_{\nu} \propto \nu^{0}$).  This stands in stark contrast to
\eventb, which was optically thick ($f_{\nu} \propto \nu^{1.3}$) up to 
high frequencies in the days and weeks following discovery
\citep{zbs+11}.  This may simply reflect physical differences in 
the jet (e.g., microphysics) or its environment (e.g., density), as is 
suggested by the different extinction properties.  However, it may
also imply a more extended (and hence long-lived) radio source, 
which would be difficult to reconcile with a tidal disruption origin.
Future high-resolution radio interferometry (i.e., VLBI) may be able
to resolve this issue.

\section{Implications}
\label{sec:implications}
The detection of a second member of this new class of relativistic
TDFs would have important implications for the rate of
these events.  Given the larger redshift of \event, a
significantly larger comoving volume (factor of 20) is accessible to detect
analogous sources.  If we adopt the local density of supermassive
black holes from \citet{tbh+07} ($\phi \approx
10^{-3}$--$10^{-2}$\,Mpc$^{-3}$) and a TDF rate per galaxy of 
$\sim R \times 10^{-5}$\,yr$^{-1}$ ($R \approx 1$--10; 
\citealt{mt99,dbe+02,esf+07,vfg+10,b11}), the all-sky rate of TDFs
within this volume is $R \times 10^{5}$\,yr$^{-1}$.  Given the
long-lived nature of the hard X-ray emission, it seems likely that the
\Swift-BAT would be capable of detecting similar outbursts within this
volume over the entire sky during the course of its 7 years of
operations to date.  Thus, if all TDFs emitted similarly bright
high-energy radiation isotropically, \Swift\ would have detected $\sim
7 R \times 10^{5}$ such outbursts to date.

Given the unique X-ray light curve, it is unlikely we have confused a
large number of these events with long-duration GRBs (or, for that
matter, other high-energy transients).  So that the
observed rate is reduced by a factor of $f_{b} \approx 10^{-6}$,
commensurate with having only seen two such events, we are
therefore left to conclude either (1) the typical beaming fraction of
these sources is extremely small ($\theta < 1^{\circ}$), or (2)
only some fraction of TDFs are capable of generating these
relativistic outflows.  Drawing an analogy with GRBs and core-collapse
supernovae, it may be that rapid rotation of the central black hole is
required to generate \event-like events.  If so, TDFs could
provide a truly unique diagnostic to constrain the spin of the central
black holes in distant galaxies. 

Late-time radio observations should be able to distinguish between
these two possibilities \citep{zbs+11,mgm12,bzp+11}.  Much as has been done with
long-duration GRBs \citep{bkf04,vkr+08}, accurate calorimetry can 
be performed in the nonrelativistic regime when the outflow is 
spherical, bypassing any concerns about geometric corrections.  
Separately, the detection of off-axis events could also help
address this fundamental issue \citep{gm11}, and such searches will be
greatly facilitated by future wide-field radio telescopes (e.g.,
LOFAR, SKA).

If, on the other hand, it is demonstrated that the emission from
\event\ was powered by an AGN (e.g., if repeated outbursts are
detected in upcoming months and years), this would add yet another
example to the already complex taxonomy of active galaxies.  The rapid
onset of accretion (when compared to the typical system lifetime,
$\Delta t_{\rm AGN} \approx 10^{7}$\,yr) would likely require some revision 
of our understanding of how material is channeled into the supermassive
black hole, and/or the physical processes responsible for launching
blazar jets.  It would also greatly complicate the search for future TDFs,
as standard spectroscopic diagnostic diagrams would appear to be
insufficient to distinguish AGNs from normal inactive galaxies. 

\acknowledgments
We wish to thank the \textit{Swift} PI N.~Gehrels and the entire
\textit{Swift} team for their work on the remarkable facilities that
enabled the discovery of this event.
We thank H.~Tananbaum for approving our \textit{Chandra} 
ToO request (ObsID 13423), and the entire \textit{Chandra} staff for
the prompt scheduling and execution of these observations.
We are grateful to G.~Fossati for providing the blazar models in
tabular form, D.~Poznanski for providing software to calculate the
host-galaxy K-corrections, and D.~Perley for assistance with the
reduction of the Keck/LRIS images.  We also acknowledge
B.~Metzger, D.~Giannos, and M. Kasliwal for valuable discussions.  
Public data from the \Swift\ data archive were used for part of this 
study.  Some of the data presented herein were obtained at the W. M. Keck 
Observatory, which is operated as a scientific partnership among the 
California Institute of Technology, the University of California and
the National Aeronautics and Space Administration; the Observatory 
was made possible by the generous financial support of the W. M. Keck 
Foundation.  S.B.C. and A.V.F. acknowledge generous financial
assistance from Gary \& Cynthia Bengier, the Richard \& Rhoda 
Goldman Fund, NASA/{\it Swift} grants NNX10AI21G and NNX12AD73G, 
the TABASGO Foundation, and NSF grant AST-0908886.

{\it Facilities:} \facility{Swift (BAT,XRT,UVOT)}, \facility{UKIRT (WFCAM)},
\facility{Keck:I (LRIS)}, \facility{Keck:II (DEIMOS)}, \facility{Max
  (GROND)}, \facility{WHT (ACAM)}


\begin{thebibliography}{82}
\expandafter\ifx\csname natexlab\endcsname\relax\def\natexlab#1{#1}\fi

\bibitem[{Abazajian {et~al.}(2009)}]{aaa+09e}
Abazajian, K.~N., {et al.} 2009, \apjs, 182, 543

\bibitem[{{Abdo} {et~al.}(2009{\natexlab{a}}){Abdo}, {Ackermann}, {Ajello},
  {Axelsson}, {Baldini}, {Ballet}, {Barbiellini}, {Bastieri}, {Baughman},
  {Bechtol}, \& et~al.}]{aaa+09h}
{Abdo}, A.~A., {et al.} 2009{\natexlab{a}}, \apj, 707, 727

\bibitem[{{Abdo} {et~al.}(2009{\natexlab{b}}){Abdo}, {Ackermann}, {Ajello},
  {Baldini}, {Ballet}, {Barbiellini}, {Bastieri}, {Bechtol}, {Bellazzini},
  {Berenji}, {Bloom}, {Bonamente}, {Borgland}, {Bregeon}, {Brez}, {Brigida},
  {Bruel}, {Burnett}, {Caliandro}, {Cameron}, {Caraveo}, {Casandjian},
  {Cecchi}, {{\c C}elik}, {Chekhtman}, {Cheung}, {Chiang}, {Ciprini}, {Claus},
  {Cohen-Tanugi}, {Conrad}, {Cutini}, {Dermer}, {de Palma}, {Silva}, {Drell},
  {Dubois}, {Dumora}, {Farnier}, {Favuzzi}, {Fegan}, {Focke}, {Foschini},
  {Frailis}, {Fukazawa}, {Fusco}, {Gargano}, {Gehrels}, {Germani}, {Giebels},
  {Giglietto}, {Giordano}, {Giroletti}, {Glanzman}, {Godfrey}, {Grenier},
  {Grove}, {Guillemot}, {Guiriec}, {Hayashida}, {Hays}, {Horan}, {Hughes},
  {J{\'o}hannesson}, {Johnson}, {Johnson}, {Kadler}, {Kamae}, {Katagiri},
  {Kataoka}, {Kerr}, {Kn{\"o}dlseder}, {Kuss}, {Lande}, {Latronico}, {Longo},
  {Loparco}, {Lott}, {Lovellette}, {Lubrano}, {Makeev}, {Mazziotta},
  {McConville}, {McEnery}, {Meurer}, {Michelson}, {Mitthumsiri}, {Mizuno},
  {Monte}, {Monzani}, {Morselli}, {Moskalenko}, {Murgia}, {Nolan}, {Norris},
  {Nuss}, {Ohsugi}, {Omodei}, {Orlando}, {Ormes}, {Pelassa}, {Pepe}, {Persic},
  {Pesce-Rollins}, {Piron}, {Porter}, {Rain{\`o}}, {Rando}, {Razzano},
  {Rochester}, {Rodriguez}, {Ryde}, {Sadrozinski}, {Sambruna}, {Sander}, {Saz
  Parkinson}, {Scargle}, {Sgr{\`o}}, {Smith}, {Spandre}, {Spinelli},
  {Strickman}, {Suson}, {Tagliaferri}, {Takahashi}, {Takahashi}, {Tanaka},
  {Thayer}, {Thayer}, {Thompson}, {Tibaldo}, {Tibolla}, {Torres}, {Tosti},
  {Tramacere}, {Uchiyama}, {Usher}, {Vasileiou}, {Vilchez}, {Vitale}, {Waite},
  {Wang}, {Winer}, {Wood}, {Ylinen}, {Ziegler}, {The Fermi/LAT Collaboration},
  {Ghisellini}, {Maraschi}, \& {Tavecchio}}]{aaa+09g}
---. 2009{\natexlab{b}}, \apjl, 707, L142

\bibitem[{{Bade} {et~al.}(1996){Bade}, {Komossa}, \& {Dahlem}}]{bkd96}
{Bade}, N., {Komossa}, S., and {Dahlem}, M. 1996, \aap, 309, L35

\bibitem[{Barthelmy {et~al.}(2005)Barthelmy, Barbier, Cummings, Fenimore,
  Gehrels, Hullinger, Krimm, Markwardt, Palmer, Parsons, Sato, Suzuki,
  Takahashi, Tashiro, \& Tueller}]{bbc+05}
Barthelmy, S.~D., {et al.} 2005, Space Science Reviews, 120, 143

\bibitem[{{Becker} {et~al.}(1995){Becker}, {White}, \& {Helfand}}]{bwh95}
{Becker}, R.~H., {White}, R.~L., and {Helfand}, D.~J. 1995, \apj, 450, 559

\bibitem[{Berger {et~al.}(2004)Berger, Kulkarni, \& Frail}]{bkf04}
Berger, E., Kulkarni, S.~R., and Frail, D.~A. 2004, \apj, 612, 966

\bibitem[{{Berger} {et~al.}(2011){Berger}, {Zauderer}, {Pooley}, {Soderberg},
  {Sari}, {Brunthaler}, \& {Bietenholz}}]{bzp+11}
{Berger}, E., {Zauderer}, A., {Pooley}, G.~G., {Soderberg}, A.~M., {Sari}, R.,
  {Brunthaler}, A., and {Bietenholz}, M.~F. 2011, arXiv e-prints
  (astro-ph/1112.1697)

\bibitem[{{Bloom} {et~al.}(2011){Bloom}, {Giannios}, {Metzger}, {Cenko},
  {Perley}, {Butler}, {Tanvir}, {Levan}, {O' Brien}, {Strubbe}, {De Colle},
  {Ramirez-Ruiz}, {Lee}, {Nayakshin}, {Quataert}, {King}, {Cucchiara},
  {Guillochon}, {Bower}, {Fruchter}, {Morgan}, \& {van der Horst}}]{bgm+11}
{Bloom}, J.~S., {et al.} 2011, Science, 333, 203

\bibitem[{{Bower}(2011)}]{b11}
{Bower}, G.~C. 2011, \apjl, 732, L12

\bibitem[{{Brandt} {et~al.}(1995){Brandt}, {Pounds}, \& {Fink}}]{bpf95}
{Brandt}, W.~N., {Pounds}, K.~A., and {Fink}, H. 1995, \mnras, 273, L47

\bibitem[{{Burrows} {et~al.}(2011){Burrows}, {Kennea}, {Ghisellini}, {Mangano},
  {Zhang}, {Page}, {Eracleous}, {Romano}, {Sakamoto}, {Falcone}, {Osborne},
  {Campana}, {Beardmore}, {Breeveld}, {Chester}, {Corbet}, {Covino},
  {Cummings}, {D'Avanzo}, {D'Elia}, {Esposito}, {Evans}, {Fugazza}, {Gelbord},
  {Hiroi}, {Holland}, {Huang}, {Im}, {Israel}, {Jeon}, {Jeon}, {Jun}, {Kawai},
  {Kim}, {Krimm}, {Marshall}, {P.~M{\'e}sz{\'a}ros}, {Negoro}, {Omodei},
  {Park}, {Perkins}, {Sugizaki}, {Sung}, {Tagliaferri}, {Troja}, {Ueda},
  {Urata}, {Usui}, {Antonelli}, {Barthelmy}, {Cusumano}, {Giommi}, {Melandri},
  {Perri}, {Racusin}, {Sbarufatti}, {Siegel}, \& {Gehrels}}]{bkg+11}
{Burrows}, D.~N., {et al.} 2011, \nat, 476, 421

\bibitem[{Burrows {et~al.}(2005)}]{bhn+05}
Burrows, D.~N., {et al.} 2005, Space Science Reviews, 120, 165

\bibitem[{Butler \& Kocevski(2007)}]{bk07}
Butler, N.~R. and Kocevski, D. 2007, \apj, 663, 407

\bibitem[{{Campana} {et~al.}(2011){Campana}, {Foschini}, {Tagliaferri},
  {Ghisellini}, \& {Covino}}]{GCN.11851}
{Campana}, S., {Foschini}, L., {Tagliaferri}, G., {Ghisellini}, G., and
  {Covino}, S. 2011, GRB Coordinates Network, 1851

\bibitem[{{Cannizzo} {et~al.}(2011){Cannizzo}, {Troja}, \& {Lodato}}]{ctl11}
{Cannizzo}, J.~K., {Troja}, E., and {Lodato}, G. 2011, \apj, 742, 32

\bibitem[{{Cenko} {et~al.}(2012){Cenko}, {Bloom}, {Kulkarni}, {Strubbe},
  {Miller}, {Butler}, {Quimby}, {Gal-Yam}, {Ofek}, {Quataert}, {Bildsten},
  {Poznanski}, {Perley}, {Morgan}, {Filippenko}, {Frail}, {Arcavi}, {Ben-Ami},
  {Cucchiara}, {Fassnacht}, {Green}, {Hook}, {Howell}, {Lagattuta}, {Law},
  {Kasliwal}, {Nugent}, {Silverman}, {Sullivan}, {Tendulkar}, \&
  {Yaron}}]{cbk+11}
{Cenko}, S.~B., {et al.} 2012, \mnras, 2203

\bibitem[{{Condon} {et~al.}(1998){Condon}, {Cotton}, {Greisen}, {Yin},
  {Perley}, {Taylor}, \& {Broderick}}]{ccg+98}
{Condon}, J.~J., {Cotton}, W.~D., {Greisen}, E.~W., {Yin}, Q.~F., {Perley},
  R.~A., {Taylor}, G.~B., and {Broderick}, J.~J. 1998, \aj, 115, 1693

\bibitem[{{Donley} {et~al.}(2002){Donley}, {Brandt}, {Eracleous}, \&
  {Boller}}]{dbe+02}
{Donley}, J.~L., {Brandt}, W.~N., {Eracleous}, M., and {Boller}, T. 2002, \aj,
  124, 1308

\bibitem[{{Esquej} {et~al.}(2007){Esquej}, {Saxton}, {Freyberg}, {Read},
  {Altieri}, {Sanchez-Portal}, \& {Hasinger}}]{esf+07}
{Esquej}, P., {Saxton}, R.~D., {Freyberg}, M.~J., {Read}, A.~M., {Altieri}, B.,
  {Sanchez-Portal}, M., and {Hasinger}, G. 2007, \aap, 462, L49

\bibitem[{{Esquej} {et~al.}(2008)}]{esk+08}
{Esquej}, P., {et al.} 2008, \aap, 489, 543

\bibitem[{{Evans} {et~al.}(2009){Evans}, {Beardmore}, {Page}, {Osborne},
  {O'Brien}, {Willingale}, {Starling}, {Burrows}, {Godet}, {Vetere}, {Racusin},
  {Goad}, {Wiersema}, {Angelini}, {Capalbi}, {Chincarini}, {Gehrels}, {Kennea},
  {Margutti}, {Morris}, {Mountford}, {Pagani}, {Perri}, {Romano}, \&
  {Tanvir}}]{ebp+09}
{Evans}, P.~A., {et al.} 2009, \mnras, 397, 1177

\bibitem[{{Faber} {et~al.}(2003){Faber}, {Phillips}, {Kibrick}, {Alcott},
  {Allen}, {Burrous}, {Cantrall}, {Clarke}, {Coil}, {Cowley}, {Davis}, {Deich},
  {Dietsch}, {Gilmore}, {Harper}, {Hilyard}, {Lewis}, {McVeigh}, {Newman},
  {Osborne}, {Schiavon}, {Stover}, {Tucker}, {Wallace}, {Wei}, {Wirth}, \&
  {Wright}}]{fpk+03}
{Faber}, S.~M., {et al.} 2003, in Society of Photo-Optical Instrumentation
  Engineers (SPIE) Conference Series, Vol. 4841, Society of Photo-Optical
  Instrumentation Engineers (SPIE) Conference Series, ed. {M.~Iye \&
  A.~F.~M.~Moorwood}, 1657--1669

\bibitem[{{Filippenko}(1982)}]{f82}
{Filippenko}, A.~V. 1982, \pasp, 94, 715

\bibitem[{{Foschini} {et~al.}(2011){Foschini}, {Ghisellini}, {Kovalev},
  {Lister}, {D'Ammando}, {Thompson}, {Tramacere}, {Angelakis}, {Donato},
  {Falcone}, {Fuhrmann}, {Hauser}, {Kovalev}, {Mannheim}, {Maraschi},
  {Max-Moerbeck}, {Nestoras}, {Pavlidou}, {Pearson}, {Pushkarev}, {Readhead},
  {Richards}, {Stevenson}, {Tagliaferri}, {Tibolla}, {Tavecchio}, \&
  {Wagner}}]{fgk+11}
{Foschini}, L., {et al.} 2011, \mnras, 413, 1671

\bibitem[{{Foschini} {et~al.}(2007){Foschini}, {Ghisellini}, {Tavecchio},
  {Treves}, {Maraschi}, {Gliozzi}, {Raiteri}, {Villata}, {Pian}, {Tagliaferri},
  {Tosti}, {Sambruna}, {Malaguti}, {Di Cocco}, \& {Giommi}}]{fgt+07}
---. 2007, \apjl, 657, L81

\bibitem[{{Fossati} {et~al.}(2008){Fossati}, {Buckley}, {Bond}, {Bradbury},
  {Carter-Lewis}, {Chow}, {Cui}, {Falcone}, {Finley}, {Gaidos}, {Grube},
  {Holder}, {Horan}, {Horns}, {Jordan}, {Kieda}, {Kildea}, {Krawczynski},
  {Krennrich}, {Lang}, {LeBohec}, {Lee}, {Moriarty}, {Ong}, {Petry}, {Quinn},
  {Sembroski}, {Wakely}, \& {Weekes}}]{fbb+08}
{Fossati}, G., {et al.} 2008, \apj, 677, 906

\bibitem[{{Fossati} {et~al.}(1998){Fossati}, {Maraschi}, {Celotti}, {Comastri},
  \& {Ghisellini}}]{fmc+98}
{Fossati}, G., {Maraschi}, L., {Celotti}, A., {Comastri}, A., and {Ghisellini},
  G. 1998, \mnras, 299, 433

\bibitem[{Gehrels {et~al.}(2004)}]{gcg+04}
Gehrels, N., {et al.} 2004, \apj, 611, 1005

\bibitem[{Gezari {et~al.}(2006)}]{gmm+06}
Gezari, S., {et al.} 2006, \apj, 653, L25

\bibitem[{Gezari {et~al.}(2008)}]{gbm+08}
---. 2008, \apj, 676, 944

\bibitem[{Gezari {et~al.}(2009)}]{ghc+09}
---. 2009, \apj, 698, 1367

\bibitem[{{Giannios} \& {Metzger}(2011)}]{gm11}
{Giannios}, D. and {Metzger}, B.~D. 2011, \mnras, 416, 2102

\bibitem[{{Goad} {et~al.}(2008){Goad}, {Tyler}, {Beardmore}, {Evans}, {Rosen},
  {Osborne}, {Starling}, {Marshall}, {Yershov}, {Burrows}, {Gehrels}, {Roming},
  {Moretti}, {Capalbi}, {Hill}, {Kennea}, {Koch}, \& {vanden Berk}}]{gtb+98}
{Goad}, M.~R., {et al.} 2008, \aap, 492, 873

\bibitem[{Greiner {et~al.}(2008)Greiner, Bornemann, Clemens, Deuter, Hasinger,
  Honsberg, Huber, Huber, Krauss, Kr{\"u}hler, Yolda{\c s}, Mayer-Hasselwander,
  Mican, Primak, Schrey, Steiner, Szokoly, Th{\"o}ne, Yolda{\c s}, Klose, Laux,
  \& Winkler}]{gbc+08}
Greiner, J., {et al.} 2008, \pasp, 120, 405

\bibitem[{{Greiner} {et~al.}(2000){Greiner}, {Schwarz}, {Zharikov}, \&
  {Orio}}]{gsz+00}
{Greiner}, J., {Schwarz}, R., {Zharikov}, S., and {Orio}, M. 2000, \aap, 362,
  L25

\bibitem[{{Grupe} {et~al.}(1995){Grupe}, {Beuermann}, {Mannheim}, {Bade},
  {Thomas}, {de Martino}, \& {Schwope}}]{gbm+95}
{Grupe}, D., {Beuermann}, K., {Mannheim}, K., {Bade}, N., {Thomas}, H., {de
  Martino}, D., and {Schwope}, A. 1995, \aap, 299, L5

\bibitem[{{G{\"u}ltekin} {et~al.}(2009){G{\"u}ltekin}, {Cackett}, {Miller}, {Di
  Matteo}, {Markoff}, \& {Richstone}}]{gcm+09}
{G{\"u}ltekin}, K., {Cackett}, E.~M., {Miller}, J.~M., {Di Matteo}, T.,
  {Markoff}, S., and {Richstone}, D.~O. 2009, \apj, 706, 404

\bibitem[{{Ho}(2009)}]{h09}
{Ho}, L.~C. 2009, \apj, 699, 626

\bibitem[{Horne(1986)}]{h86}
Horne, K. 1986, \pasp, 98, 609

\bibitem[{Kalberla {et~al.}(2005)Kalberla, Burton, Hartmann, Arnal, Bajaja,
  Morras, \& P{\"o}ppel}]{kbh+05}
Kalberla, P. M.~W., Burton, W.~B., Hartmann, D., Arnal, E.~M., Bajaja, E.,
  Morras, R., and P{\"o}ppel, W. G.~L. 2005, \aap, 440, 775

\bibitem[{{Kann} {et~al.}(2010){Kann}, {Klose}, {Zhang}, {Malesani}, {Nakar},
  {Pozanenko}, {Wilson}, {Butler}, {Jakobsson}, {Schulze}, {Andreev},
  {Antonelli}, {Bikmaev}, {Biryukov}, {B{\"o}ttcher}, {Burenin}, {Castro
  Cer{\'o}n}, {Castro-Tirado}, {Chincarini}, {Cobb}, {Covino}, {D'Avanzo},
  {D'Elia}, {Della Valle}, {de Ugarte Postigo}, {Efimov}, {Ferrero}, {Fugazza},
  {Fynbo}, {G{\aa}lfalk}, {Grundahl}, {Gorosabel}, {Gupta}, {Guziy}, {Hafizov},
  {Hjorth}, {Holhjem}, {Ibrahimov}, {Im}, {Israel}, {Je{\'l}inek}, {Jensen},
  {Karimov}, {Khamitov}, {Kizilo{\v g}lu}, {Klunko}, {Kub{\'a}nek}, {Kutyrev},
  {Laursen}, {Levan}, {Mannucci}, {Martin}, {Mescheryakov}, {Mirabal},
  {Norris}, {Ovaldsen}, {Paraficz}, {Pavlenko}, {Piranomonte}, {Rossi},
  {Rumyantsev}, {Salinas}, {Sergeev}, {Sharapov}, {Sollerman}, {Stecklum},
  {Stella}, {Tagliaferri}, {Tanvir}, {Telting}, {Testa}, {Updike}, {Volnova},
  {Watson}, {Wiersema}, \& {Xu}}]{kkz+07}
{Kann}, D.~A., {et al.} 2010, \apj, 720, 1513

\bibitem[{{Kelson}(2003)}]{k03}
{Kelson}, D.~D. 2003, \pasp, 115, 688

\bibitem[{Kinney {et~al.}(1996)Kinney, Calzetti, Bohlin, McQuade,
  Storchi-Bergmann, \& Schmitt}]{kcb+96}
Kinney, A.~L., Calzetti, D., Bohlin, R.~C., McQuade, K., Storchi-Bergmann, T.,
  and Schmitt, H.~R. 1996, \apj, 467, 38

\bibitem[{{Komossa} \& {Greiner}(1999)}]{kg99}
{Komossa}, S. and {Greiner}, J. 1999, \aap, 349, L45

\bibitem[{{Krimm} {et~al.}(2011){Krimm}, {Kennea}, {Holland}, {Barthelmy},
  {Baumgartner}, {Cummings}, {Fenimore}, {Gehrels}, {Markwardt}, {Palmer},
  {Sakamoto}, {Skinner}, {Stamatikos}, {Tueller}, \& {Ukwatta}}]{ATEL.3384}
{Krimm}, H.~A., {et al.} 2011, The Astronomer's Telegram, 3384

\bibitem[{{Krolik} \& {Piran}(2011)}]{kp11}
{Krolik}, J.~H. and {Piran}, T. 2011, \apj, 743, 134

\bibitem[{Kulkarni {et~al.}(1998)Kulkarni, Frail, Wieringa, Ekers, Sadler,
  Wark, Higdon, Phinney, \& Bloom}]{kfw+98}
Kulkarni, S.~R., {et al.} 1998, Nature, 395, 663

\bibitem[{{Lauer} {et~al.}(2007)}]{lfr+07}
{Lauer}, T.~R., {et al.} 2007, \apj, 662, 808

\bibitem[{{Levan} {et~al.}(2011){Levan}, {Tanvir}, {Cenko}, {Perley},
  {Wiersema}, {Bloom}, {Fruchter}, {de Ugarte Postigo}, {O'Brien}, {Butler},
  {van der Horst}, {Leloudas}, {Morgan}, {Misra}, {Bower}, {Farihi},
  {Tunnicliffe}, {Modjaz}, {Silverman}, {Hjorth}, {Thoene}, {Cucchiara},
  {Castro Ceron}, {Castro-Tirado}, {Arnold}, {Bremer}, {Brodie}, {Carroll},
  {Cooper}, {Curran}, {Cutri}, {Ehle}, {Forbes}, {Fynbo}, {Gorosabel},
  {Graham}, {Guizy}, {Hoffman}, {Jakobsson}, {Kamble}, {Kerr}, {Kasliwal},
  {Kouveliotou}, {Kocesvki}, {Law}, {Nugent}, {Ofek}, {Poznanski}, {Quimby},
  {Rol}, {Romanowsky}, {Sanchez-Ramirez}, {Schulze}, {Singh}, {Starling},
  {Strom}, {Wheatley}, {Wijers}, {Winters}, {Wold}, \& {Xu}}]{ltc+11}
{Levan}, A.~J., {et al.} 2011, Science, 333, 199

\bibitem[{{Magorrian} \& {Tremaine}(1999)}]{mt99}
{Magorrian}, J. and {Tremaine}, S. 1999, \mnras, 309, 447

\bibitem[{{Magorrian} {et~al.}(1998)}]{mtr98}
{Magorrian}, J., {et al.} 1998, \aj, 115, 2285

\bibitem[{Maksym {et~al.}(2010)Maksym, Ulmer, \& Eracleous}]{mue10}
Maksym, W.~P., Ulmer, M.~P., and Eracleous, M. 2010, \apj, 722, 1035

\bibitem[{{Metzger} {et~al.}(2012){Metzger}, {Giannios}, \& {Mimica}}]{mgm12}
{Metzger}, B.~D., {Giannios}, D., and {Mimica}, P. 2012, \mnras, 2207

\bibitem[{{Miller} \& {G{\"u}ltekin}(2011)}]{mg11}
{Miller}, J.~M. and {G{\"u}ltekin}, K. 2011, \apjl, 738, L13

\bibitem[{{Murray} {et~al.}(1997){Murray}, {Chappell}, {Kenter}, {Kobayashi},
  {Kraft}, {Meehan}, {Zombeck}, {Fraser}, {Pearson}, {Lees}, {Brunton},
  {Pearce}, {Barbera}, {Collura}, \& {Serio}}]{mck+97}
{Murray}, S.~S., {et al.} 1997, in Society of Photo-Optical Instrumentation
  Engineers (SPIE) Conference Series, Vol. 3114, Society of Photo-Optical
  Instrumentation Engineers (SPIE) Conference Series, ed. {O.~H.~Siegmund \&
  M.~A.~Gummin}, 11--25

\bibitem[{Oke \& Gunn(1983)}]{og83}
Oke, J.~B. and Gunn, J.~E. 1983, \apj, 266, 713

\bibitem[{Oke {et~al.}(1995)}]{occ+95}
Oke, J.~B., {et al.} 1995, \pasp, 107, 375

\bibitem[{{Perley} {et~al.}(2011){Perley}, {Chandler}, {Butler}, \&
  {Wrobel}}]{pcb+11}
{Perley}, R.~A., {Chandler}, C.~J., {Butler}, B.~J., and {Wrobel}, J.~M. 2011,
  \apjl, 739, L1

\bibitem[{{Pian} {et~al.}(2007){Pian}, {Romano}, {Treves}, {Ghisellini},
  {Covino}, {Cucchiara}, {Dolcini}, {Tagliaferri}, {Markwardt}, {Campana},
  {Chincarini}, {Gehrels}, {Giommi}, {Maraschi}, {Vergani}, {Zerbi},
  {Molinari}, {Testa}, {Tosti}, {Vitali}, {Antonelli}, {Conconi}, {Malaspina},
  {Nicastro}, {Palazzi}, {Meurs}, \& {Norci}}]{prt+07}
{Pian}, E., {et al.} 2007, \apj, 664, 106

\bibitem[{{Pian} {et~al.}(1998){Pian}, {Vacanti}, {Tagliaferri}, {Ghisellini},
  {Maraschi}, {Treves}, {Urry}, {Fiore}, {Giommi}, {Palazzi}, {Chiappetti}, \&
  {Sambruna}}]{pvt+98}
---. 1998, \apjl, 492, L17

\bibitem[{{Pineau} {et~al.}(2011){Pineau}, {Motch}, {Carrera}, {Della Ceca},
  {Derri{\`e}re}, {Michel}, {Schwope}, \& {Watson}}]{pmc+11}
{Pineau}, F.-X., {Motch}, C., {Carrera}, F., {Della Ceca}, R., {Derri{\`e}re},
  S., {Michel}, L., {Schwope}, A., and {Watson}, M.~G. 2011, \aap, 527, A126

\bibitem[{{Poole} {et~al.}(2008)}]{pbp+08}
{Poole}, T.~S., {et al.} 2008, \mnras, 383, 627

\bibitem[{{Rau} {et~al.}(2011{\natexlab{a}}){Rau}, {Greiner}, \&
  {Olivares}}]{ATEL.3390}
{Rau}, A., {Greiner}, J., and {Olivares}, F. 2011{\natexlab{a}}, The
  Astronomer's Telegram, 3390

\bibitem[{{Rau} {et~al.}(2011{\natexlab{b}}){Rau}, {Greiner}, {Schady},
  {Olivares}, {Krimm}, \& {Holland}}]{ATEL.3425}
{Rau}, A., {Greiner}, J., {Schady}, P., {Olivares}, F., {Krimm}, H., and
  {Holland}, S. 2011{\natexlab{b}}, The Astronomer's Telegram, 3425

\bibitem[{{Readhead}(1994)}]{r94}
{Readhead}, A.~C.~S. 1994, \apj, 426, 51

\bibitem[{{Rees}(1988)}]{r88}
{Rees}, M.~J. 1988, \nat, 333, 523

\bibitem[{{Remillard} \& {McClintock}(2006)}]{rm06}
{Remillard}, R.~A. and {McClintock}, J.~E. 2006, \araa, 44, 49

\bibitem[{Renzini {et~al.}(1995)Renzini, Greggio, di~Serego~Alighieri,
  Cappellari, Burstein, \& Bertola}]{rgd+95}
Renzini, A., Greggio, L., di~Serego~Alighieri, S., Cappellari, M., Burstein,
  D., and Bertola, F. 1995, Nature, 378, 39

\bibitem[{Roming {et~al.}(2005)}]{rkm+05}
Roming, P. W.~A., {et al.} 2005, Space Science Reviews, 120, 95

\bibitem[{Schlegel {et~al.}(1998)Schlegel, Finkbeiner, \& Davis}]{sfd98}
Schlegel, D.~J., Finkbeiner, D.~P., and Davis, M. 1998, \apj, 500, 525

\bibitem[{Skrutskie {et~al.}(2006)}]{scs+06}
Skrutskie, M.~F., {et al.} 2006, \aj, 131, 1163

\bibitem[{{Socrates}(2011)}]{s11}
{Socrates}, A. 2011, arXiv e-prints (astro-ph/1105.2557)

\bibitem[{Spergel {et~al.}(2007)}]{sbd+07}
Spergel, D.~N., {et al.} 2007, \apjs, 170, 377

\bibitem[{{Tundo} {et~al.}(2007){Tundo}, {Bernardi}, {Hyde}, {Sheth}, \&
  {Pizzella}}]{tbh+07}
{Tundo}, E., {Bernardi}, M., {Hyde}, J.~B., {Sheth}, R.~K., and {Pizzella}, A.
  2007, \apj, 663, 53

\bibitem[{{Urry} \& {Padovani}(1995)}]{up95}
{Urry}, C.~M. and {Padovani}, P. 1995, \pasp, 107, 803

\bibitem[{van~der Horst {et~al.}(2008)van~der Horst, Kamble, Resmi, Wijers,
  Bhattacharya, Scheers, Rol, Strom, Kouveliotou, Oosterloo, \&
  Ishwara-Chandra}]{vkr+08}
van~der Horst, A.~J., {et al.} 2008, \aap, 480, 35

\bibitem[{{van Velzen} {et~al.}(2011){van Velzen}, {Farrar}, {Gezari},
  {Morrell}, {Zaritsky}, {{\"O}stman}, {Smith}, {Gelfand}, \& {Drake}}]{vfg+10}
{van Velzen}, S., {et al.} 2011, \apj, 741, 73

\bibitem[{{V{\'e}ron-Cetty} \& {V{\'e}ron}(2010)}]{vv10}
{V{\'e}ron-Cetty}, M.-P. and {V{\'e}ron}, P. 2010, \aap, 518, A10

\bibitem[{Voges {et~al.}(1999)}]{vab+99}
Voges, W., {et al.} 1999, \aap, 349, 389

\bibitem[{Wehrle {et~al.}(1998)}]{wpu+98}
Wehrle, A.~E., {et al.} 1998, \apj, 497, 178

\bibitem[{{Zauderer} {et~al.}(2011){Zauderer}, {Berger}, {Soderberg}, {Loeb},
  {Narayan}, {Frail}, {Petitpas}, {Brunthaler}, {Chornock}, {Carpenter},
  {Pooley}, {Mooley}, {Kulkarni}, {Margutti}, {Fox}, {Nakar}, {Patel},
  {Volgenau}, {Culverhouse}, {Bietenholz}, {Rupen}, {Max-Moerbeck}, {Readhead},
  {Richards}, {Shepherd}, {Storm}, \& {Hull}}]{zbs+11}
{Zauderer}, B.~A., {et al.} 2011, \nat, 476, 425

\end{thebibliography}


\clearpage
\begin{deluxetable}{lccccc}
\tabletypesize{\scriptsize}
\tablecaption{UV/Optical/NIR Observations of \event}
\tablewidth{0pt}
\tablehead{
\colhead{Date\tablenotemark{a}} & \colhead{Telescope/Instrument} & 
\colhead{Filter} &  \colhead{Exposure Time} & 
\colhead{Magnitude\tablenotemark{b}} \\
\colhead{(2011 UT)} & & & \colhead{(s)} &
}
\startdata
May 27.91 & UVOT & $b$ & 2927.3 & $> 21.50$ \\
May 28.38 & GROND & $J$ & 720.0 & $> 19.6$ \\
May 28.38 & GROND & $H$ & 720.0 & $> 19.0$ \\
May 28.38 & GROND & $K$ & 720.0 & $> 17.3$ \\
May 29.41 & GROND & \gp & 3600.0 & $22.55 \pm 0.06$ \\
May 29.41 & GROND & \rp & 3600.0 & $22.66 \pm 0.07$\\
May 29.41 & GROND & \ip & 3600.0 & $22.89 \pm 0.18$\\
May 29.41 & GROND & \zp & 3600.0 & $22.91 \pm 0.27$\\
May 29.41 & GROND & $J$ & 3600.0 & $> 21.0$\\
May 29.41 & GROND & $H$ & 3600.0 & $> 20.5$\\
May 29.41 & GROND & $K$ & 3600.0 & $> 18.0$\\
May 30.58 & UVOT & $uvw1$ & 2580.1 & $> 22.80$ \\
Jun 2.41 & UVOT & $uvm2$ & 2889.8 & $> 23.07$ \\
Jun 3.40 & GROND & \gp & 1440.0 & $22.57 \pm 0.06$ \\
Jun 3.40 & GROND & \rp & 1440.0 & $22.68 \pm 0.09$ \\
Jun 3.40 & GROND & \ip & 1440.0 & $22.81 \pm 0.15$ \\
Jun 3.40 & GROND & \zp & 1440.0 & $23.20 \pm 0.34$ \\
Jun 3.40 & GROND & $J$ & 1440.0 & $> 21.4$ \\
Jun 3.40 & GROND & $H$ & 1440.0 & $> 20.9$ \\
Jun 3.40 & GROND & $K$ & 1440.0 & $> 18.8$ \\
Jun 5.88 & UVOT & $uvw2$ & 3059.2 & $> 23.37$ \\
Jun 8.57 & UVOT & $u$ & 2258.0 & $21.51 \pm 0.29$ \\
Jun 10.26 & GROND & \gp & 9600.0 & $22.40 \pm 0.05$ \\
Jun 10.26 & GROND & \rp & 9600.0 & $22.62 \pm 0.06$ \\
Jun 10.26 & GROND & \ip & 9600.0 & $22.98 \pm 0.11$ \\
Jun 10.26 & GROND & \zp & 9600.0 & $23.17 \pm 0.20$ \\
Jun 10.26 & GROND & $J$ & 9600.0 & $> 21.9$ \\
Jun 10.26 & GROND & $H$ & 9600.0 & $> 21.2$ \\
Jun 10.26 & GROND & $K$ & 9600.0 & $> 19.3$ \\
Jun 14.04 & UVOT & $uvm2$ & 3113.3 & $> 23.13$ \\
Jun 17.11 & UVOT & $uvw2$ & 2823.8 & $> 23.43$ \\
Jun 18.51 & WFCAM & $H$ & 2400.0 & $> 20.8$ \\
Jun 18.53 & WFCAM & $K$ & 2400.0 & $> 20.9$ \\
Jun 23.54 & UVOT & $uvw1$ & 2434.2 & $> 22.92$ \\
Jun 26.55 & UVOT & $uvm2$ & 2468.0 & $> 23.12$ \\
Jun 29.45 & Keck/LRIS & \gp & 300.0 & $22.75 \pm 0.04$ \\
Jun 29.45 & Keck/LRIS & $R$ & 280.0 & $22.82 \pm 0.04$ \\
Jul 1.76 & UVOT & $uvw1$ & 3050.1 & $> 23.02$ \\
Jul 6.46 & UVOT & $u$ & 2402.1 & $22.09 \pm 0.29$ \\
Jul 7.00 & UVOT & $uvw2$ & 430.1 & $> 22.21$ \\
Jul 9.15 & WHT & \rp & 300.0 & $22.96 \pm 0.04$ \\
Jul 9.17 & WHT & \gp & 300.0 & $23.02 \pm 0.05$ \\
Jul 11.53 & UVOT & $uvw2$ & 2271.5 & $> 23.36$ \\
\enddata
\tablenotetext{a}{UT at beginning of exposure.}
\tablenotetext{b}{Reported magnitudes have not been corrected for 
  Galactic extinction ($E(B-V) = 0.095$\,mag; \citealt{sfd98}).  Upper
  limits represent $3\sigma$ uncertainties.}
\label{tab:opt}
\end{deluxetable}

\end{document}